# Agile Approach for IT Forensics Management

Matthias Schopp
*Research Institute CODE*
*Bundeswehr University Munich*
matthias.schopp@unibw.de

Peter Hillmann
*Research Institute CODE*
*Bundeswehr University Munich*
peter.hillmann@unibw.de

## Abstract

*The forensic investigation of cyber attacks and IT incidents is becoming increasingly difficult due to increasing complexity and intensify networking. Especially with Advanced Attacks (AT) like the increasing Advanced Persistent Threats an agile approach is indispensable. Several systems are involved in an attack (multi-host attacks). Current forensic models and procedures show considerable deficits in the process of analyzing such attacks. For this purpose, this paper presents the novel flower model, which uses agile methods and forms a new forensic management approach. In this way, the growing challenges of ATs are met. In the forensic investigation of such attacks, big data problems have to be solved due to the amount of data that needs to be analyzed. The proposed model meets this requirement by precisely defining the questions that need to be answered in an early state and collecting only the evidence usable in court proceedings that is needed to answer these questions. Additionally, the novel flower model for AT is presented that meets the different phases of an investigation process.*

## 1. Introduction

In the last year's industry, governments and institutions had to face more and more complex and enduring cyber-attacks, such as ThyssenKrupp [1] or RUAG [2]. Mandiant reported attacks on a wide spectrum of industries in 2015 [3]. The median number of days a threat group was active in a network before detection has been 205 days in year 2014, where the longest presence was 2982 days. Such advanced attacks (AT) are long-running, complex and often have the goal to steal or destroy data. The know-how of the attackers is high and therefore the attacks are highly sophisticated and heterogeneous. Furthermore, the attack tool support for the attackers are getting more and more professionalized, whereas the development of forensic software is second-order driven.

## 2. Problem Description

Digital forensic is defined as analytical and investigative techniques used for the preservation, identification, extraction, documentation, analysis, and interpretation of computer media (e.g. data on SSD/HDD/DVDs, log data on host systems or log data for network activities). When it comes to ATs the amount of computer media that has to be analyzed in a forensic process is increasing. This is partly due to its long duration and persistence. ATs have a higher sophistication level as basic or moderate attacks and are more difficult to clarify. In particular through the cautious and covert approach, whereby slow progress from the attack side is consciously accepted in order not to be discovered.

Forensic investigations of ATs therefore need to adapt to these characteristics. Classical forensic approaches that focus on individual systems (classical post-mortem forensic) and that collect all data in a first step and then try to extract hypotheses from this data is no longer expedient (shotgun forensic vs. sniper forensic [4]). This is because on the one hand not only individual systems are affected but also the communication between them (network forensic). On the other hand the amount of data becomes bigger and therefore big data analysis approaches are required that define questions asked in a first manner and then select the data that can answer them.

Therefore a forensic approach for ATs is needed that is agile and takes network forensic as well as big data into account. Due to the complexity of ATs the new model also aims to fit the different steps a high sophisticated intruder makes during the attack.

## 3. Related Work

For a digital forensic model, it is important to guide the investigation process in a continuously transparent way. When it comes to ATs the model also needs to be flexible because of the heterogeneity of ATs. In the past, several forensic models have been presented.

Du et al. [5] describe three phases in the current evolution of digital forensic process models. From 2000 to 2010 general models have been defined, such as McKemmish's SAP model (secure, analyze, and present) in [6] or more detailed ones such as the extended model of cyber-crime investigation [7] that enhances the model of Casey [8] that provides a scientific basis for the digital forensic process. These early frameworks have a linear procedure in common and they do not take ATs and big data problems into account.

In the second phase, these models have been refined and the newer approaches focus on particular steps in the investigation process. In [9] Agarwal et al. present a model that focuses on computer fraud and cybercrimes. None of them do treat ATs or focus on big data issues.

The third mentioned phase in [5] describes the recent research in this field. New technologies are taken into account such as cloud computing and internet of things (IoT). Some research has been done when it comes to big data and forensic. In [10] a framework for data reduction and data mining is presented that is not suggested to replace full analysis but serves to provide a rapid triage. They conclude that this approach is time consuming but has its advantages in the field of topic browsing.

All of the presented work are collecting data first and then try to reduce the amount of data or extract the relevant evidence instead of defining the questions asked in a first step. None of the existing models meets the requirements for an agile sniper forensic model. Although these aforesaid approaches can be used as guide in our new model.

## 4. Research Questions

The forensic process faces many challenges when it comes to ATs. The characteristics of these kind of attacks create a big data scenario that has to be addressed in research. The heterogeneity of ATs also has to be taken into account for a forensic model. Therefore the following research questions are expected to be answered.

*1) What does the procedure look like when it comes to the forensic investigation of advanced attacks having regard to the heterogeneity of the attacks?*

*2) How can the relevant data for the investigation process be identified and forensically usable extracted having regard to the relevant questions and considering the big data problems?*

*3) How do host and network sensors (e.g. intrusion detection, network taps, etc.) have to be placed and forensically secured in the network to capture the needed data for the investigation process?*

## 5. Agile Sniper Forensics for Advanced Attacks Approach

The novel approach focuses on the forensic investigation of advanced attacks and meets the modern challenges of an agile sniper approach. It consists of three parts: The two sub processes *Investigation* and *Documentation* as well as the novel flower model.

### 5.1 Flower Model

The flower model illustrates the attack procedure presented in Figure 1 taking Mandiant's attack lifecycle [3] and Hutchins et al. intrusion kill chain [11] into account. The model separates the attack in two perspectives, *Attacker* and *Target*. This is because from a forensic process' point of view in the step *Initial Information Gathering* an attacker uses public available information and legitimate tools and therefore this is hard to detect and to prove as part of the attack. Nevertheless, a regularly self-check to delete sensible public data or generatable information higher the attack difficulty for an intruder and should be part of the restoration process after the forensic investigation.

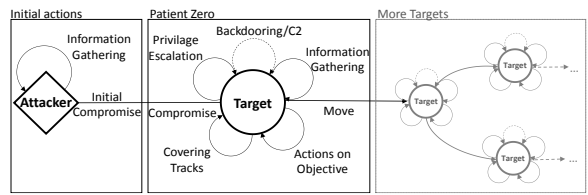

**Figure 1 Flower Model for attack step illustration**

The initial information gathering results in the initial compromise of the first target system. Each compromised system during the attack is represented as a flower (right hand side of the figure) with different steps as the flowers leaves. These represent the other phases of the attack such as *escalating privileges* after a successful compromise, *maintaining access* and control over the target through *backdoors* and the connection to a *command and control server (C2)*, *information gathering* on the host and/or network, e.g. through several scan techniques, *actions on objective* such as data exfiltration and *covering tracks*. The arrow compound the result either of the attacker's i*nitial compromise* or the action *move* from the perspective of a *Target*. ADs usually include multiple systems that are compromised due to lateral movement steps. These steps are illustrated by the *move* action in the Flower Model where the first compromised system's (patient zero) C*ompromise* action results from the *initial compromise* of the attacker. The model also takes into account that there can be more than one initial compromised system, which leads to multiple *initial compromise* actions, as well as more than one *move* action.

The *Target*'s get connected to an attack graph that represents the involved compromised systems and illustrates the attack chain.

## 5.2 Invesitgation and documentation method (IDM)

The digital forensic process makes use of the flower model to comprehend the AT. Figure 2 shows this agile sniper forensic approach.

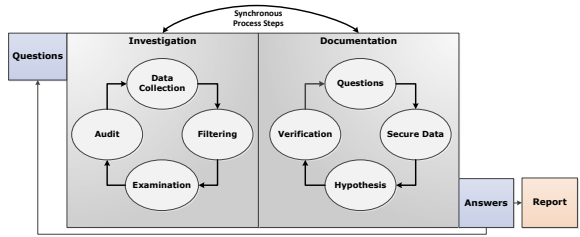

**Figure 2. Agile Sniper Forensic Model**

It is divided into two sub processes, *Investigation* and *Documentation* where these two processes exist simultaneously. The global management process starts with defining the general *questions* that need to be answered to forensically reproduce the AT. For each found compromised system, the questions related to it are defined and the sub processes are done. According to Greiner [12] examples for such questions are:

- How did the attacker(s) get onto the system?
- What did the attacker(s) do on the system and what did they steal?
- How did the attacker(s) escape / move laterally?
- How did the attacker(s) get the stolen data off the system?

The *Investigation* sub process starts with collecting the data that is needed to answer the defined questions. These data can be rough categorized in host, network and misc. Sources of information are usually log files and temporary files as well as status information like timestamps, user rights, and processes. When it comes to AT the amount of data that needs to be analyzed is big. Therefore in our approach these data sources that are relevant to the defined questions need to be defined and only that data is collected. This subset of data than is filtered to get the specific data for the separate questions. It is examined so hypotheses can be developed that answer the defined questions. The results are then checked with the found data.

In parallel the *Documentation* sub process documents the investigation steps. The questions for extracting the necessary data as well as the steps to collect it are recorded. Therefore an official and common agreed process is mandatory, which is integrity securing. The collected data is stored with checksums and signatures in a matter that its integrity is provable. During the examination in the *Investigation* process the developed hypotheses also need to be documented. The last step is the reporting of the audit actions to verify the hypotheses.

If more data is needed to answer the questions, other filters need to be applied, hypotheses are refuted, or new hypotheses are developed that need to be proven, these sub processes will be repeated until the answers to the questions are proven with the data.

The whole process is also repeated until the initial attack vector for the compromised system is found and its origin (system or attacker) is found. The model is designed to adapt the questions depending on the found evidence on a system. This is to be agile on the steps taken by the attacker. Each compromised system is treated this way during the digital forensic investigation. The documentation builds the separate flowers from the flower model with its leaves and connects them so the complete attack is forensically reproducible in the report.

## 6. Evaluation

We analysed 21 technical reports on ADs ([2], [13], [14], [15], [16], [17], [18], [19], [20], [21], [22], [23], [24], [25], [26], [27], [28], [29], [30], [31], [32]). These reports analysed one or more attacks that used similar types of malwares and tools per report. To show that an agile approach is needed we examined the amount of targets involved in such an attack and the steps taken by the attackers.

The reports show that in all 21 cases multiple targets were part of the attacks. The six types of actions on a target – the leaves of the flower model – were also present on the analysed reports. The actions and tools used by the attackers were heterogeneous and targeted on their objectives. Therefore the forensic analysis process needs to be agile to the attack. The presented approach will be evaluated in more detail in a practical manner in future work.

## 7. Conclusion and Next Steps

In this proposal, the need for a digital forensic management approach for advanced attacks was pointed out. With the agile sniper forensic model processes were presented as a first step towards such a management approach. For the ease of reproducibility also the flower model was introduced that can be used for documentation purposes.

In the next steps, this forensic approach has to be refined and tested in a practical manner. Therefore a specific tool support will be developed. It also has to be analyzed what techniques can be used to identify and extract relevant data and also where to place sensors to capture the relevant data.